\documentclass[%
 aip,
 jmp,%
 amsmath,amssymb,
 reprint,%
]{revtex4-1}

\usepackage{graphicx}
\usepackage{dcolumn}
\usepackage{bm}

\begin{document}

\preprint{AIP/123-QED}

\title{Appearance of Large Amplitude Current Dynamics as a Result of Bragg Reflection in Carbon Nanotubes}

\thanks{Corresponding author. Tel.:+233 042 33837 Email: profsymensah@yahoo.co.uk}

\author{S. S. Abukari}%
\affiliation{Department of Physics, Laser and Fibre Optics Centre, University of Cape Coast, Cape Coast, Ghana.
}%

\author{S. Y. Mensah}
\affiliation{Department of Physics, Laser and Fibre Optics Centre, University of Cape Coast, Cape Coast, Ghana.
}%

\author{N. G. Mensah}
\affiliation{Department of Mathematics, University of Cape Coast, Cape Coast, Ghana.
}%

\author{K. W. Adu}
\affiliation{Department of Physics, The Pennsylvania State University Altoona College, Altoona, Pennsylvania 16601, USA. Research Institute, The Pennsylvania State University, University Park, PA 16802, USA
}%

\author{M. Rabiu}
 \affiliation{Department of Applied Physics, Faculty of Applied Sciences, University for Development Studies, Navrongo Campus, Ghana.}

\author{K. A. Dompreh}
\affiliation{Department of Physics, Laser and Fibre Optics Centre, University of Cape Coast, Cape Coast, Ghana.
}%

\author{A. Twum}
\affiliation{Department of Physics, Laser and Fibre Optics Centre, University of Cape Coast, Cape Coast, Ghana.
}%

\date{\today}

\begin{abstract}
We report on theoretical analysis of  large amplitude current dynamics due to Bragg reflections in carbon nanotubes exposed to an external electric field. Using the kinetic equation with constant relaxation time,  an analytical expression for the current density is obtained. Our results suggest that Bloch gain exists up to frequencies on the order of the Bloch frequency. We noted that due to the high density of states of conduction electrons in metallic carbon nanotubes and the specific dispersion law inherent in hexagonal crystalline structure result in a uniquely high frequency gain than the corresponding values for semiconducting ones. We suggest that this phenomenon can be used for domainless multiplication of the frequency of an electromagnetic signal at room temperature.
%
\end{abstract}



\maketitle

\section{Introduction}
The nonlinear dynamics of electrons in carbon nanotubes (CNs) under the action of an external electric field has been the subject of intense research \cite{1, 4}. Tans {\it et al.} \cite{1} and Bezryadin {\it et al.} \cite{2} have measured the  current-voltage characteristics for single-wall CNs at low temperatures, i.e when $k_BT < \epsilon_C$, $\Delta \epsilon$. Where $k_B$ is Boltzmann constant, $T$ is the temperature, $\epsilon_C$ is the charging energy. The energy level spacing $\Delta \epsilon$ is given by $\Delta \epsilon = \pi \hbar v_F/L$ where $v_F$ is the Fermi velocity and L is the carbon nanotube length \cite{4}. At this low temperatures the current is produced by the electrons tunneling through the CNs in the presence of the Coulomb blockade induced by long-range Coulomb interactions \cite{4}. Nonlinear coherent transport through doped nanotube junctions was considered in Ref. \cite{3} and it also showed the possibility of negative differential conductivity (NDC) for tunelling electrons. Using semiclassical Boltzmann’s equation, Maksimenko {\it et al.}l \cite{4} found that the dc current- voltage characteristics of CNs biased by a constant electric field along the axis of undoped CNTs at room temperatures ( $k_BT > \epsilon_C$, $\Delta\epsilon$ ) showed NDC up to frequencies on the order of the Bloch frequency. Here the current density is produced by conduction electrons with energies below the energy of the interband transitions and move in the crystalline field like free quasiparticles, with a modified dispersion law allowing the use of quasiclassical approach to describe the electron motion. Bloch oscillations have been observed in semiconductor superlattices \cite{5, 6}. These Bloch oscillations are caused by Bragg reflection of electrons at the edges of the Brillouin regions and the occurrence of Bloch oscillations are concluded from the observation of NDC.

Like Maksimenko {\it et al.} \cite{4}, quantum phenomena like interband transitions, quantum mechanical corrections to intraband motion as well as Coulomb interactions will be excluded from this report. Our report will be quasiclassical with an extension to the high-frequency electric field in CNs by following the approach of Ktitorov {\it et al.}  \cite{7}.

\section{Theoretical model}
Following \cite{4}, the high-frequency differential conductivity is derived starting with the Boltzmann kinetic equation with \cite{5}. For simplicity, we shall ignore the difference between the energy - momentum relaxation rates and assume a common relaxation time $\tau$ for both processes. The equations of the symmetric $f_s$ and antisymmetric $f_a$ parts of the distribution functions are 
\begin{equation}
	\frac{\partial f_s}{\partial t} + eE(t)\frac{\partial f_a}{\partial k_x} = - \frac{f_s - f_0(p)}{\tau} \label{eq:one}
\end{equation}
\begin{equation}
	\frac{\partial f_a}{\partial t} + eE(t)\frac{\partial f_s}{\partial k_x} = - \frac{f_a}{\tau} \label{eq:two}
\end{equation}
Writing the distribution functions in Fourier series as
\begin{equation}
	f_a^1 = \Delta p_{\varphi} \sum_{s=1}^{n}\delta(p_{\varphi} - s\Delta p_{\varphi})\sum_{r\neq 0} f_{rs}e^{iarp_z}\Phi_a^1 \label{eq:three}
\end{equation}
\begin{equation}
	f_s^1 = \Delta p_{\varphi} \sum_{s=1}^{n}\delta(p_{\varphi} - s\Delta p_{\varphi})\sum_{r\neq 0} f_{rs}e^{iarp_z}\Phi_s^1 \label{eq:four}
\end{equation}
\begin{equation}
	f_0(p) = \Delta p_{\varphi} \sum_{s=1}^{n}\delta(p_{\varphi} - s\Delta p_{\varphi})\sum_{r\neq 0} f_{rs}e^{iarp_z} \label{eq:five}
\end{equation}
where $e$ is the electron charge, $f_0(p)$ is the equilibrium distribution function, $\delta(x)$ is the Dirac delta function, $r$ is summation over the stark component, $ f_{rs}$ is the coefficient of the Fourier series and $\Phi_s^1$ and $\Phi_a^1$ are the factors by which the Fourier transform of the symmetric $f_s^1$ and antisymmetric $f_a^1$ parts of the distribution functions are different from the equilibrium distribution function $f_0(p, t)$, and $\tau_e = \tau_p = \tau$ is the relaxation time. The electric field $E$  is applied along CNs axis.  The equilibrium distribution function can be expanded in the analogous series with the coefficients $f_{rs}$ expressed as
 \begin{equation}
 	  f_{rs} = \frac{a}{2\pi}\int_0^{\frac{2\pi}{a}} \frac{e^{iarp_z}}{1 + exp(\epsilon_s(p_z))/k_BT)} \label{eq:six}
\end{equation}    

The investigation is done within the semiclassical approximation in which conduction electrons with energy below the energy of the interband transitions move in the crystalline lattice like free quasi-particles with dispersion law extracted from quantum theory \cite{4}. Taking into account the hexagonal crystalline structure of a rolled graphene in a form of CNT and using the tight binding approximation, the energy dispersion for zigzag and armchair CNTs for which the  Fermi energy $\epsilon_F=0$, are  expressed as in Eqns \eqref{eq:seven} and \eqref{eq:eight}, respectively \cite{4, 8, 9}
 \begin{eqnarray}
	\epsilon_s(s\Delta p_{\varphi}, p_z) &\equiv& \epsilon_s(p_z)\nonumber\\
	 &=& \pm\gamma_0\Big[ 1 + 4 cos (ap_z) cos\Big(\frac{a}{\sqrt{3}}s\Delta p_\varphi\Big) \ldots\nonumber \\
	 && + 4cos^2\Big( \frac{a}{\sqrt{3}}s\Delta p_{\varphi}\Big)\Big]^{1/2}\label{eq:seven}
\end{eqnarray}   
\begin{eqnarray}
	\epsilon_s(s\Delta p_{\varphi}, p_z) &\equiv& \epsilon_s(p_z)\nonumber\\
	 &=& \pm\gamma_0\Big[ 1 + 4 cos (as\Delta p_{\varphi}) cos\Big(\frac{a}{\sqrt{3}} p_z\Big) \ldots\nonumber\\
	 && + 4 cos^2\Big(\frac{a}{\sqrt{3}} p_z\Big)\Big]^{1/2}\label{eq:eight}
\end{eqnarray}    
where $\gamma_0 \sim 3.0$ eV is the overlapping integral, $p_z$ is the axial component of quasi-momentum, $\Delta p_{\varphi}$ is transverse quasi-momentum  level spacing and s is an integer. The expression for a in Eqns \eqref{eq:seven} and \eqref{eq:eight} is given as  $a=3b⁄2\hbar$, $b=0.142$ nm is the C-C bond length. The – and + signs correspond to the valence and conduction bands respectively. Due to the transverse quantization of the quasi-momentum, its transverse component can take n discrete values, $p_{\varphi} = s\Delta p_{\varphi} = \pi\sqrt{3}s⁄an$   ($s=1, \ldots, n$). Unlike transverse quasi-momentum $p_{\varphi}$, the axial quasi momentum   is assumed to vary continuously within the range $0 \leq p_z \leq 2\pi⁄a$, which corresponds to the model of infinitely long  of CNT ($L = \infty$). 

The quasiclassical velocity $v_z(p_z, s∆p_{\phi})$ of an electron moving along the CNT axis is given by the expression
\begin{equation}
	v_z (p_z, s∆p_z ) = \gamma_0 \sum_{r \neq 0} \frac{\partial (\epsilon_{rs} e^{iarp_z} )}{\partial p_z } = \gamma_0 \sum_{r  \neq 0} iar \epsilon_{rs} e^{iarp_z}     \label{eq:nine}
\end{equation}

Substituting Eqns \eqref{eq:three}, \eqref{eq:four} and \eqref{eq:five} into Eqns \eqref{eq:one} and \eqref{eq:two} and solving with the perturbations $f_s = f_s^0 + f_s^1 e^{-i\omega t}$, $f_a = f_a^0 + f_a^1 e^{-i\omega t}$ and $E = E_0 + E_1 e^{-i \omega t}$, we obtain the surface current density defined by
\[
	j_z = \frac{2e}{(2\pi \hbar)^2}\int\int f(p,t)  v_z (p) d^2 p
\]
or
\begin{equation}
	j_z = \frac{2e}{(2\pi \hbar)^2}\sum_{s=1}^{n}\int_0^{\frac{2\pi}{a}} f\big(p_z, s\Delta p_{\varphi}, \Phi(t) \big)  dp_z \label{eq:ten}
\end{equation}
as
\begin{eqnarray}
	\sigma_z (\omega) &=& \sigma_0 \frac{4e^2\gamma_0\sqrt{3}}{n\hbar^2} \sum_{r=1}r\left[ \frac{1 - i\omega\tau^2 - \omega_B^2 \tau^2}{(\omega_B^2 - \omega^2) \tau^2 + 1 - 2i\omega\tau} \right] \nonumber \\
		&& \times \sum_{s=1}^n f_{rs} \epsilon_{rs} \label{eq:eleven}
\end{eqnarray}
\begin{equation}
               \sigma_0 = \frac{\sigma_0 (0)}{(\omega_B^2\tau^2 + 1) }  \label{eq:twelve}
\end{equation}

Here, $j_0 = \frac{4 e^2 \gamma_0 \sqrt{3}}{n\hbar^2}$. $\omega_B = eaE/\sqrt{3}$ for armchair and $\omega_B = eaE_0$ for zigzag CNs.
\begin{figure}[ht!]
\includegraphics[scale=.8]{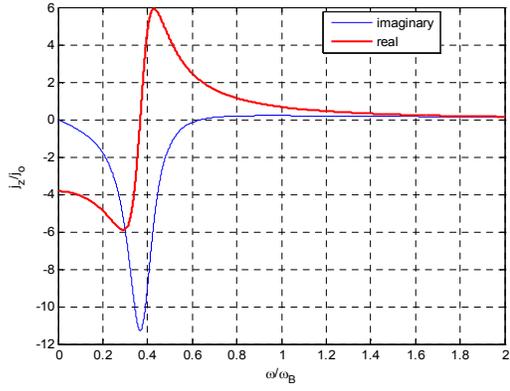}
\caption{Real (red) and imaginary (blue) parts of conductivity for armchair CNs for $\omega/\omega_B = 10$.}\label{fig:one}
\end{figure}

\begin{figure}[ht!]
\includegraphics[scale=.8]{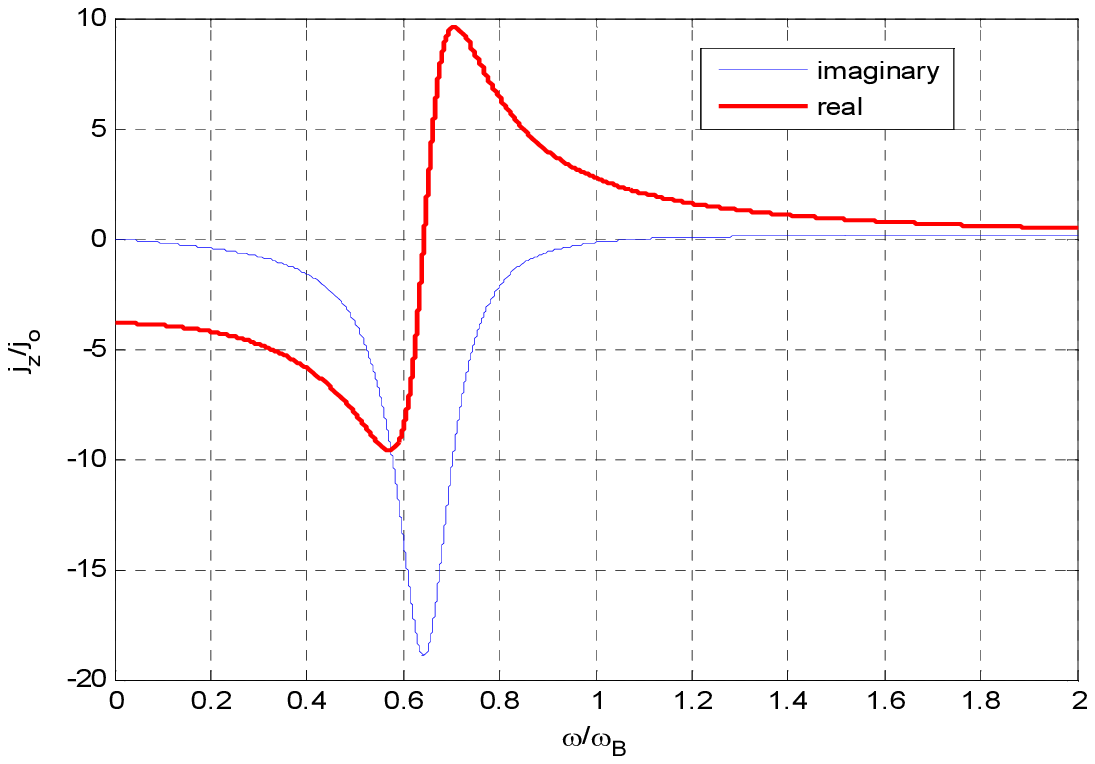}
\caption{Real (red) and imaginary (blue) parts of conductivity for zigzag CNs for $\omega/\omega_B = 10$.}\label{fig:two}
\end{figure}

\section{Results, Discussion and Conclusion}
Using quasiclassical approach, we consider an influence of Bragg reflections on dynamics of an electron transport in a CNs simultaneously exposed to both constant (dc) and ac electric fields. An analytical expression for the current density is obtained. The nonlinearity is analyzed basically on the dependence of the current density on the frequency.

In Figs.\ref{fig:one} and \ref{fig:two}, we show the plot of the real and imaginary parts of the normalized current density ($j_z/j_0$) as a function of dimensionless frequency ($\omega/\omega_B$) for a-SWCNT and z-SWCNT, respectively. In all two systems, we observed that $Re(\sigma_z)$ is negative for $\omega < 1\tau  \sqrt{(\omega_B \tau )^4 - 1) /( (\omega_B\tau)^2 - 1)}$ which is indicative for gain at these frequencies. 

In conclusion,    we have large amplitude current dynamics due to Bragg reflections of carbon nanotubes effect in undoped CNs. We have obtained an expression for the current density-electric field characteristics for CNs in presence of dc-ac fields. We noted that the high density of states of conduction electrons in metallic carbon nanotubes and the specific dispersion law inherent in hexagonal crystalline structure result in a uniquely high gain at these frequencies than the corresponding values for semiconducting ones.


\end{document}